\documentclass[fleqn,12pt]{wlscirep}

\usepackage{accents}

\usepackage{graphicx}
\usepackage{caption}
\usepackage{subcaption}

\newcommand{\all}[2]{\,\begin{align}
#1 
\label{#2}
\end{align}
}

\title{An Epidemiological  Mechanistic Model \\  of Box Office Dynamics }

\author[1,*]{Naghmeh Momeni}
\author[2] {Amir Tohidi Kalorazi}
 \author[1]{Michael  Rabbat}
\author[3,4]{Babak Fotouhi}
\affil[1]{Department of Electrical and Computer Engineering, McGill University, Montreal, Quebec, Canada }
\affil[2]{Department of Electrical  Engineering, Sharif University of Technology, Tehran, Iran }
\affil[3]{Program for Evolutionary Dynamics, Harvard University, Cambridge, MA, USA}
\affil[4]{Institute for Quantitative Social Sciences, Harvard University, Cambridge, MA, USA}

\affil[*]{naghmeh.momenitaramsari@mail.mcgilll.ca}

\begin{abstract}
In this paper we propose a mechanistic model that links micro social interactions to macro observables in the case of diffusion of 
 film-going decisions. 
We  devise   a  generalized  epidemic model to  capture  the temporal evolution of  box  office  revenues.
The model adds social influence and memory effects to conventional epidemic models. Fitting the model to a temporal data set containing domestic weekly revenue of  the     5000  top  US-grossing films of all time, we find that a two-parameter model can capture a remarkable  portion of the observed variance in the data. 
 Using the  distribution of the  estimated parameters for different genres, we  then present a predictive model which provides reasonable  a-priori estimates 
 of future sale as a function of time. 
\end{abstract}
\begin{document}

\flushbottom
\maketitle
%
%
\thispagestyle{empty}









\section*{Introduction}

 Understanding human collective behavior is a much-studied topic across a diverse array of disciplines. 
There are retrospective approaches and predictive approaches. In the former, the aim  is to understand  how (and why)  certain  collective social movements emerged. Examples include  revolutions~\cite{skocpol1994social,skocpol1979states}, sudden emergence of  radical nationwide attitudes~\cite{arendt1963eichmann,neumann2013secret},  financial crises~\cite{stiglitz2010freefall,rajan2011fault}, and even more mundane  subjects such as fashion  trends and fads~\cite{davis1994fashion,bikhchandani1992theory}. The level of complexity of  collective social  problems is too high to allow simple answers, which is manifest in the fact that for many of the said problems, there is still no convergent framework, and consensus is rarely reached on how (and especially why) certain social phenomena emerged. The predictive approach is more common in disciplines such as public policy,  finance, economics, and marketing. The aim is to   forecast  how people would react to certain policies, technologies, products, etc., with reasonable degree of accuracy.  Consumer behavior, word-of-mouth effects and herd behaviors  are central topics  in behavioral finance and marketing research~\cite{banerjee1992simple,herd2,peter1999consumer,solomon2014consumer}. The main reason is clear; in a socio-economic system whose centerpiece is the free market,   investors   naturally like to  increase investment efficiency by foreseeing the future, even partially.  Though progress is being made,  the status-quo in many domains is still far from near-certain prediction. 
The popular adage ``Half the money I spend on advertising is wasted; the trouble is I don't know which half'',  often attributed to the  marketing pioneer John Wanamaker~\cite{watts2011everything}, is still relevant to the present-day state of knowledge. The popular accounts on this  challenge of prediction   is also ample. One example is the rejection of   \emph{Harry Potter}  by 12 publishers before finally getting published, which  summarizes the current ability of even experts  in predicting  product success and consumer behavior.


In this paper, we  focus on  a particular example of consumption, and seek to describe it with  as simple a model  as possible. We focus on films.  Forecasting box office performance of films is a well-researched topic~\cite{neelamegham1999bayesian,ainslie2005modeling,elberse2003demand,eliashberg2006motion,krider1998competitive,einav2007seasonality,ravid1999information}. 
  In the  post-Internet era (which has transformed the typical size and scope of available data),  more and more attention is being paid   to search counts, weblog content, and social media trends, to understand and predict consumer behavior in every marketing context, including  film revenues. 
Most  of these studies aim to predict future revenues based on pre/upon-release data, such as pre-release critiques~\cite{joshi2010movie,smith461movie},  Twitter mentions~\cite{asur2010predicting}, YouTube views~\cite{apala2013prediction}, pre-release news mentions~\cite{zhang2009improving}, Wikipedia page hits and edits~\cite{Taha}, and  search engine query data~\cite{goel2010predicting}.
   Although studio research teams are advancing knowledge on what makes a film popular, the current understanding of consumer behavior in this context and the current ability to predict future box office success is still far from certainty.

 In contrast to the  approaches of the   studies mentioned above, we take a mechanistic, bottom-up approach in this paper to study the evolution of film popularity. 
We are interested in modeling  the micro interpersonal  mechanisms that drive this dynamics, and seek to capture the emergence of macro observable outcomes (which has the advantage of data availability). 
 Films constitute a special kind of commodity, with idiosyncratic features in terms of diffusion. 
 In the present stage of capitalist economy, most products are subject to a constant influx:  frequently,  new versions of products (e.g., cell phones) arrive, and  people update their possessions at varying speeds (some must have the most recent iPhone as soon  as possible, some wait  longer, some are pushed via planned obsolescence, etc.). This is more or less true for a considerable share of products with use: cell phones,  TVs,  video game consoles, laptops, and tablets, to name a few examples. 
 Films, however,  have different dynamics (let us emphasize that we focus only on box-office sales; we are not considering subsequent revenues via DVD release and other royalties). 
  Almost all people go to the cinema  \emph{together}. That is, film-going is a joint activity. So the social component  that drives the film-going decision is strong. 
  There is also a time factor: films do not stay on the screen indefinitely and are sequentially  replaced by new ones.  In this paper we seek to  construct a  minimal model that captures these features. 
  
We devise a basic model of  social influence derived from conventional epidemic models (see~\cite{pastor2015epidemic} for a thorough review of these models) with a twist.  By adding memory effects to a basic epidemic-type model, we obtain a two-parameter model which is analytically tractable. We use data on the weekly domestic (in the USA) sales  of 5000   recent films to fit model parameters, and show that the model can reliably capture the temporal evolution of the  box-office revenue. 
Although our main aim is to shed light on the micro social mechanisms, and not prediction,  we show that as a byproduct of the analysis, the results can also have considerable  prediction-oriented utility with remarkable accuracy, which we do analyze but not as the central theme of the paper.

\section*{Results}

 \subsection*{model}  
 We denote an individual who has seen the film with $I$ (for \emph{infected}) and one who has not seen the film by $S$ (for \emph{susceptible}). Social influence  results from a person observing friends, family, peers, etc., having seen a film. If the proportion of social ties  in one's proximity  having seen the film increases, the person is more likely to catch up with the trend.  
  For a given social tie between an $S$ individual and an $I$ individual, we denote the `transmission' rate by $\beta$. That is, for a small time increment $dt$, the probability that the $S$ individual gets infected as a result of this social tie is $\beta dt$. 
  It is  worth noting that in actuality, there are two distinct compartments that both exert influence on an $S$ individual to see the film. 
  One consists of social ties who have already seen the film, and the other consists of those who have decided to see the film but have not   done so yet. Since people rarely go to the theater alone, the latter group is expected to have considerable contribution in the social influence on an individual.  Because model simplicity is our
  first priority in this study, we approximate the situation by assuming that the $I$ compartment consists of both of the   subpopulations mentioned above, and we assume that $\beta$ is an average transmission coefficient for both types. In other words, we assume that `having decided to see the film' and `having seen the film' are interchangeable, and we are ignoring the cases where a decision is made but eventually  not followed through on.

  Another central element of the dynamics is memory. There is a constant inflow of new films being released. Human cognitive capacity is limited. 
  The `buzz' created by a newly-released film inevitably decreases as time advances. 
   A simple choice to model human memory is an  exponential function $e^{-B t}$, where $B>0$ is  a constant~\cite{sudman1973effects,lu1992behavioral}. The larger the value of $B$ is, the shorter the lifetime of the film will be in the collective memory of the society. 
   Also, we expect film lifetime (the number of weeks theaters screen it) to decrease with $B$.  The reason is that increasing $B$ makes the film being forgotten more quickly, and  theaters only keep screening a   films until the generated revenue matches the costs.
  
  Let us denote the cumulative box office revenue at time $t$ by $G(t)$, and let $\rho_t$  denote  the proportion of individuals  in the $I$ compartment at time $t$. 
  The temporal evolution of  $\rho(t)$ under the mean-field regime    is given by 
  \all{
  \dot{\rho}_t=   (1-\rho_t) \rho_t \beta \langle k \rangle e^{-Bt} 
  ,}{rhodot}
 where $\langle k \rangle$ is the average number of social contacts of individuals. Denoting $\beta \langle k \rangle / B$ by $A$, 
   and noting that $G(t)$ is proportional to $\rho_t$, 
 we obtain the following expression for $G(t)$:
 \all{
 G(t)= G(0)  \displaystyle  e^{\displaystyle A (1-e^{-Bt})}
 .}{Gt}
  
  It is worth noting that $A$ models social influence, that is, stronger $A$ means stronger effect from the social surroundings on the decision of filmgoers. The value of $A$ can increase in two ways: increase in the connectivity of people (through $\langle k \rangle$), or increase in the per-link transmission rate.
  
 An alternative approach to modeling social influence is taking into account group pressure, instead of individual influence.  For example, in the 
 conventional voter model (which is the basic   model of opinion dynamics with a vast literature of studies and extensions~\cite{voter}), the probability of switching to one state from another is determined by the  respective proportions of the states at the individual's vicinity. 
 In the epidemic approach, we assumed that each social tie exerts an influence to an individual which is independent of the number of social ties that person has. An $S$ person with three $I$ friends receives  the same amount of influence whether this person has 20 friends or only has those three friends. The voter model takes a different approach, and considers  proportions instead of absolute numbers. This means that  the $S$ person with three $I$ friends will receive an influence proportional to $3/20$ if the person has 20 friends, and proportional  to $3/3$ if the person has three friends.  An alternative voter-type model can be devised for the problem at hand which leads to the same end result under the mean field approximation. If individual $x$  with state $S$ has $k_x$ friends and $I_x$ of them have state $I$, then the probability 
  that individual $x$ will change its state to $I$ is given by $\alpha dt  ({I_x}/{k_x}) e^{-Bt}$. The memory factor is identical to the previous scenario, because memory-wise there is no difference between the two cases. The factor $\alpha$ is just the rate of strategy revision, that is, how frequently on average do individuals consider updating their state. In the methods section we show that in this case, 
  The temporal evolution of  $\rho(t)$ under the mean-field regime    is given by 
  \all{
  \dot{\rho}_t=   (1-\rho_t) \rho_t \alpha  e^{-Bt} 
  ,}{rhodot2}
 which is identical to~\eqref{rhodot}, with only the name of the coefficient changed. The parametric  form of  equation for $G(t)$ will be identical to the one obtained above.

 \subsection*{Parameter Estimation}
 
 Let us define $z(t) \stackrel{\text{def}}{=} \log \big[ G(t)/G(0) \big] $.
 To estimate $A$ and $B$ in~\eqref{Gt} for a given film, what we essentially need to do is to fit the following model to empirical time-series data of film sales: $
z(t) = A (1-e^{-Bt})$. Minimizing the sum of squared errors (denoted by $E$)  via  conventional   algorithms such as gradient descent or Newton's method is then straightforward, especially because  the gradient and the Hessian are very simple to calculate:

 \all{
\resizebox{.95\linewidth}{!}{$
 \nabla E= 
 \displaystyle
 \sum_t
\begin{bmatrix}
 -2 \left(1-e^{-B t}\right) \left(z_t-A \left(1-e^{-B t}\right)\right)
\\
 -2 A t e^{-B t} \left(z_t-A \left(1-e^{-B t}\right)\right) 
\end{bmatrix}
 ,~~
 H= 
  \displaystyle
  \sum_t
 \begin{bmatrix}
 2 \left(-1+e^{-B t}\right)^2 
 & 
 e^{-2 B t} \left(2 t e^{B t}   (2 A-z_t)-4 A t\right) 
 \\
 e^{-2 B t} \left(2 t e^{B t}   (2 A-z_t)-4 A t\right)
  & 
 2 A t^2 \left(2 A+e^{B t} (z_t-A)\right) e^{-2 B t} 
  \\
 \end{bmatrix}
 $}
}{nababa}

\begin{figure}[!h]
        \centering
        \begin{subfigure}[b]{0.3 \columnwidth}
                \includegraphics[width=\columnwidth]{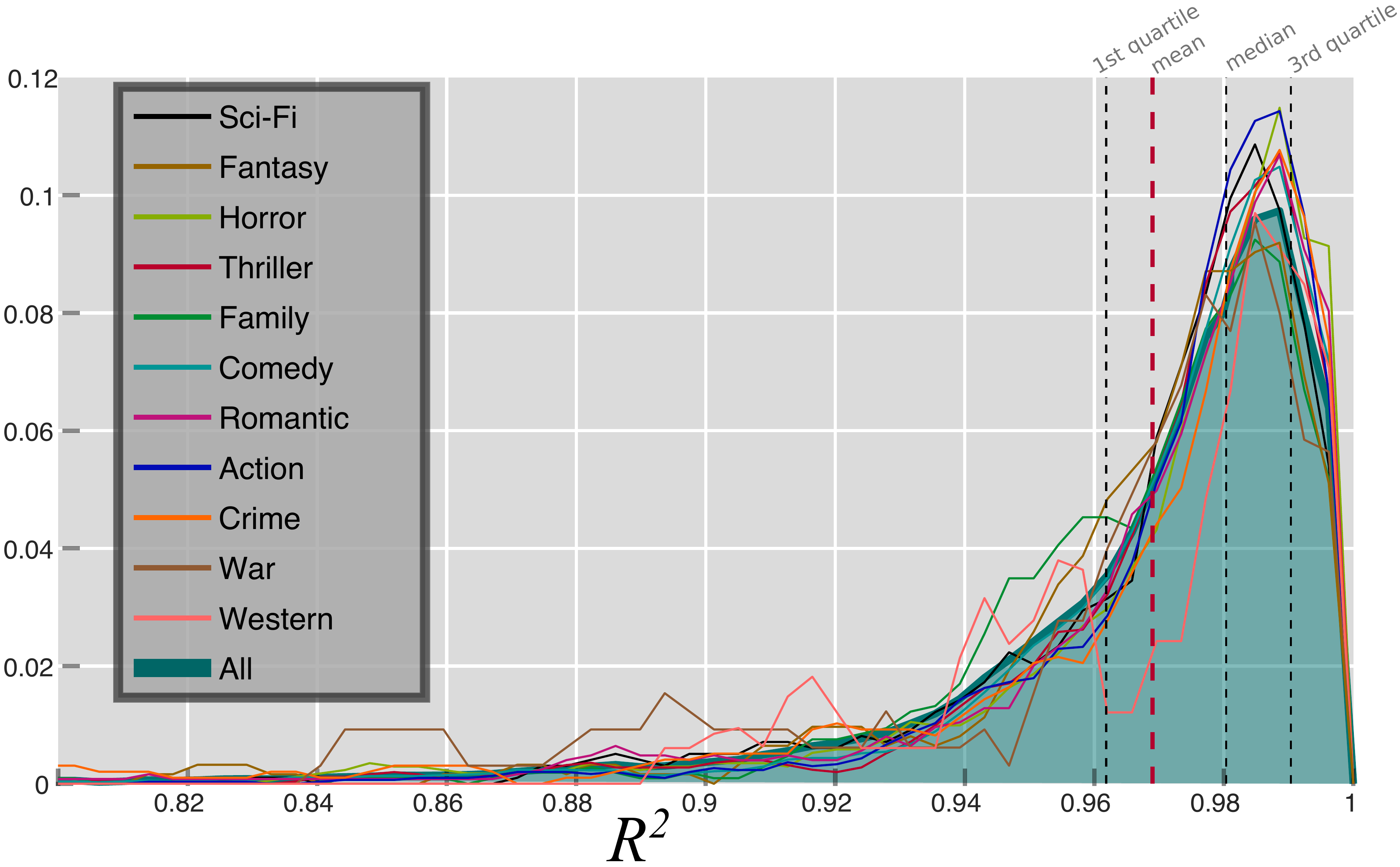}
                \caption{ }
                \label{R2_in_genres}
        \end{subfigure}%
        ~ 
            \begin{subfigure}[b]{0.3 \columnwidth}
                \includegraphics[width=\columnwidth]{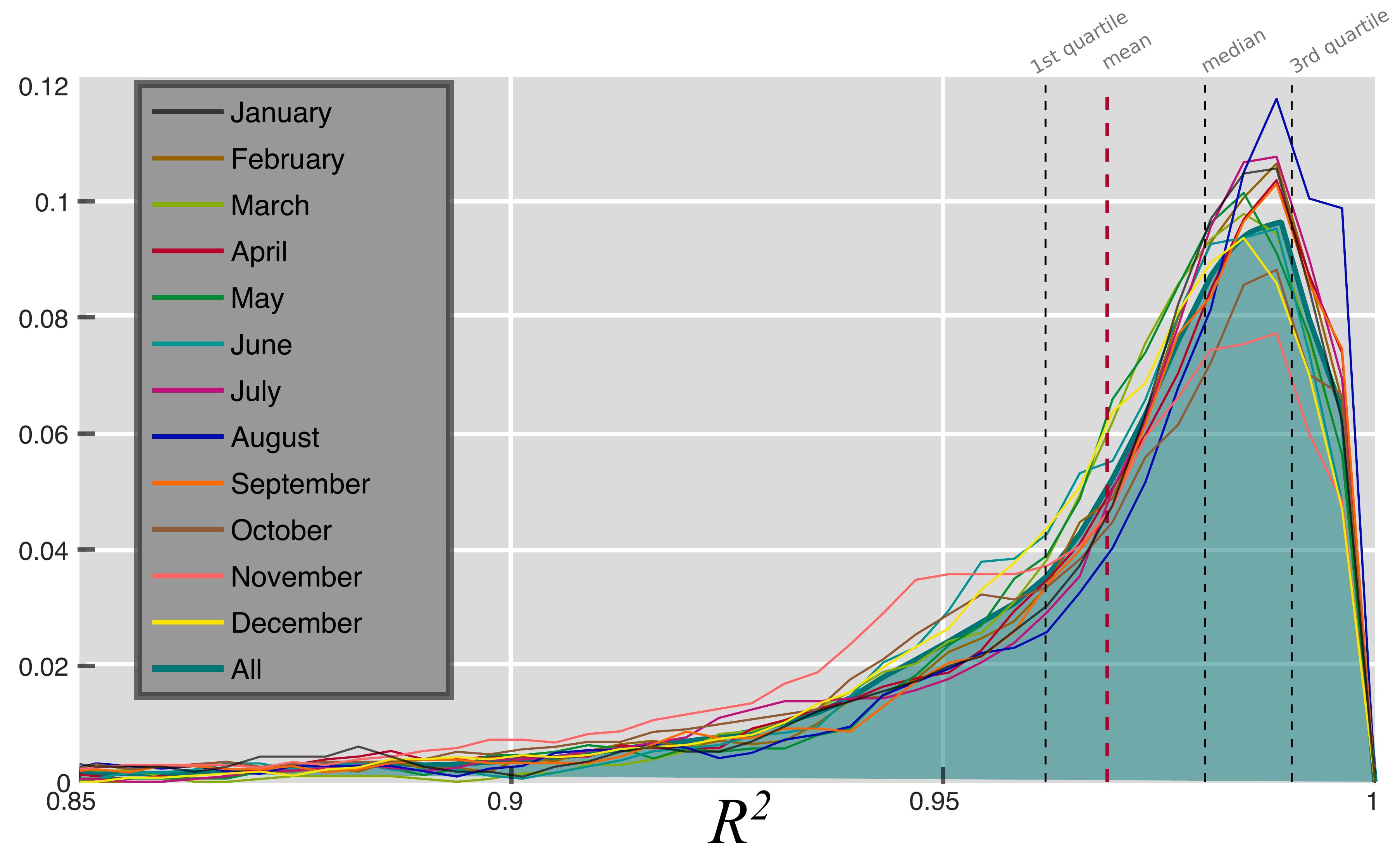}
                \caption{ }
                \label{R2_in_months}
        \end{subfigure}   
~
         \begin{subfigure}[b]{0.3 \columnwidth}
                \includegraphics[width=\columnwidth]{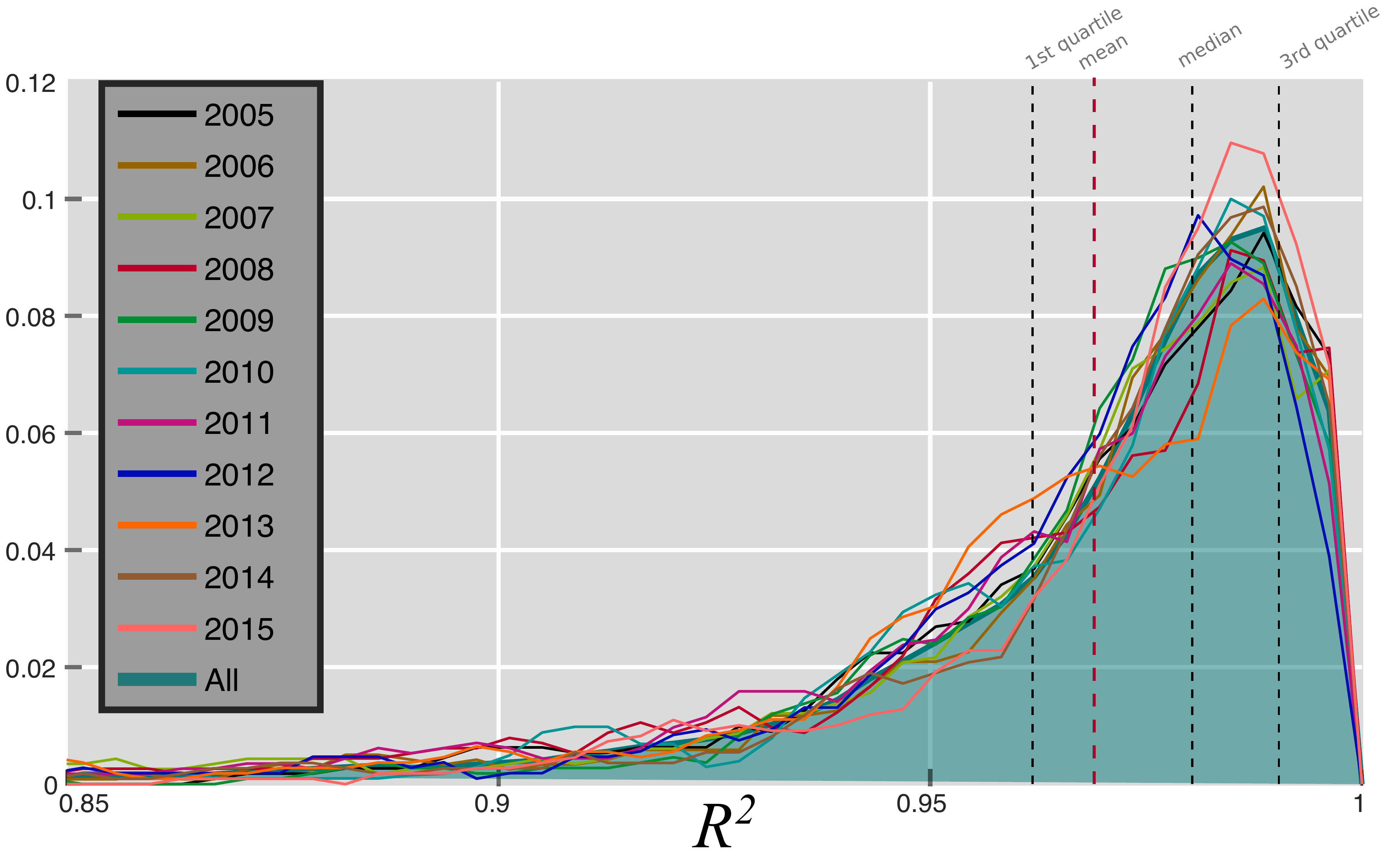}
                \caption{ }
                \label{R2_in_years}
        \end{subfigure}%
        \caption{ The histogram of adjusted $R^2$ for films as categorized by (a) genre, (b) month of release, (c) year of release.   }\label{R2}
\end{figure}

\begin{figure}[!h]
        \centering
        \begin{subfigure}[b]{0.3 \columnwidth}
                \includegraphics[width=\columnwidth ,height=4cm]{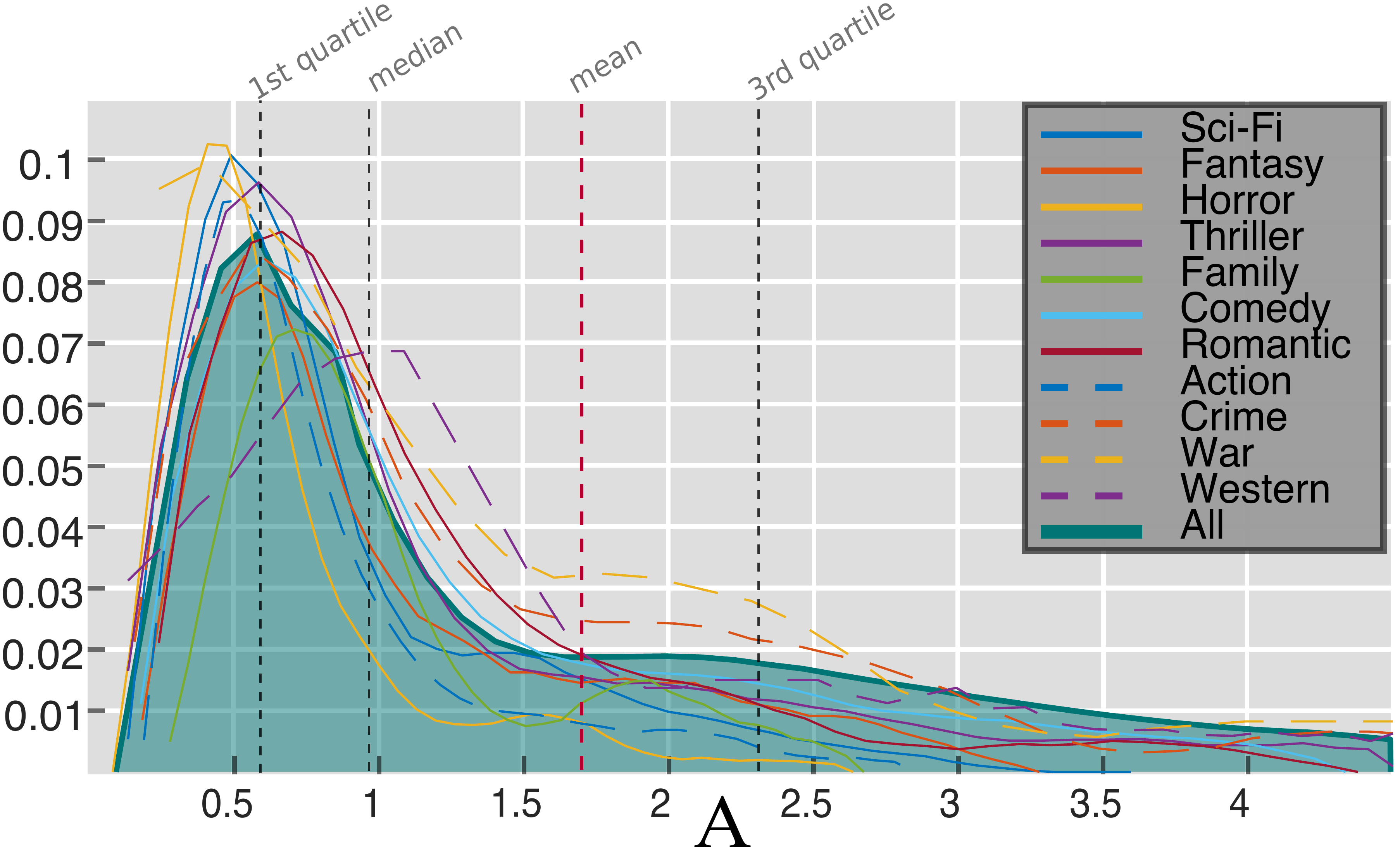}
                \caption{ }
                \label{A_hist_in_genres}
        \end{subfigure}%
        ~ 
            \begin{subfigure}[b]{0.3 \columnwidth}
                \includegraphics[width=\columnwidth,height=4cm ]{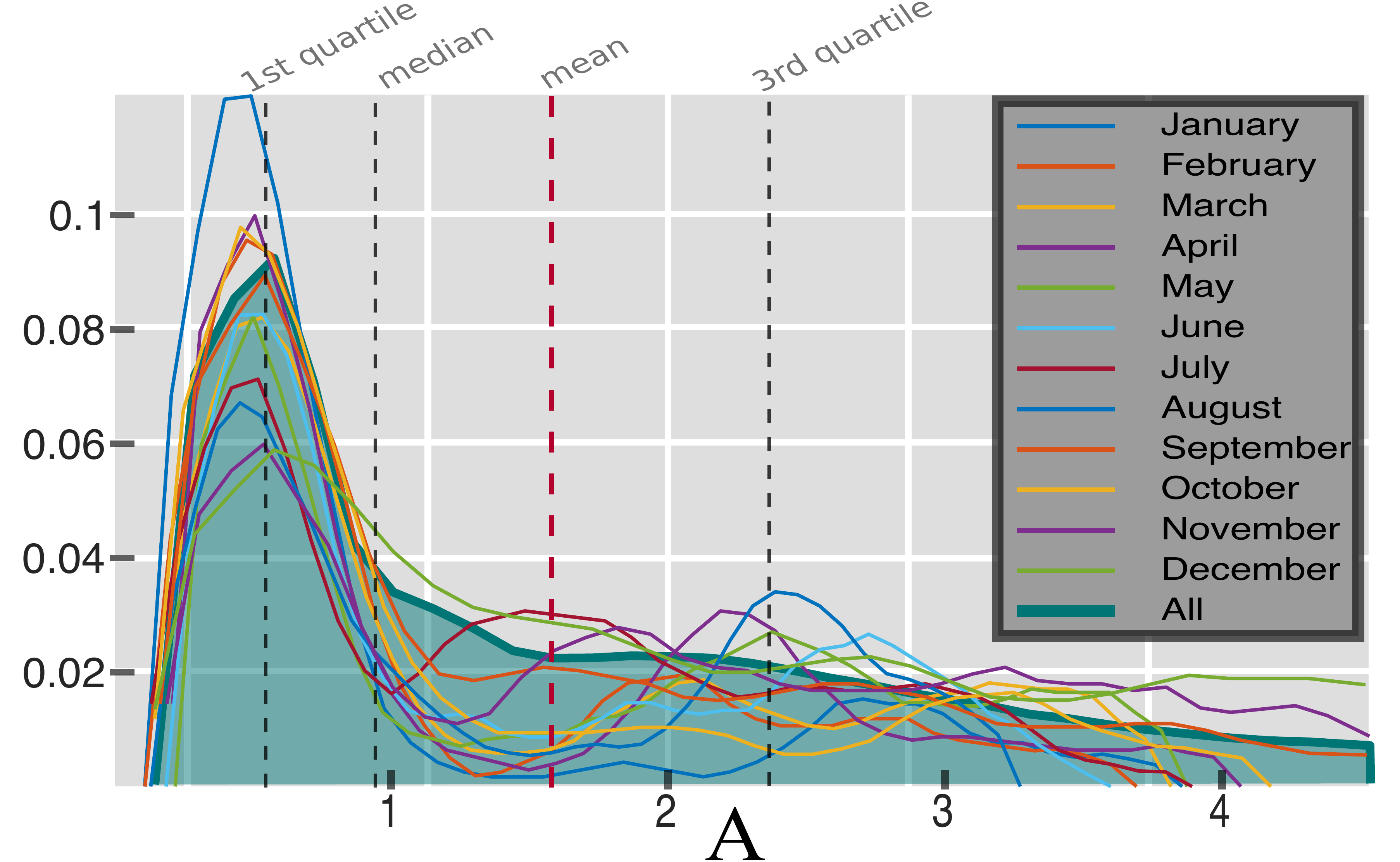}
                \caption{ }
                \label{A_hist_in_months}
        \end{subfigure}   
~
         \begin{subfigure}[b]{0.3 \columnwidth}
                \includegraphics[width=\columnwidth ,height=4cm]{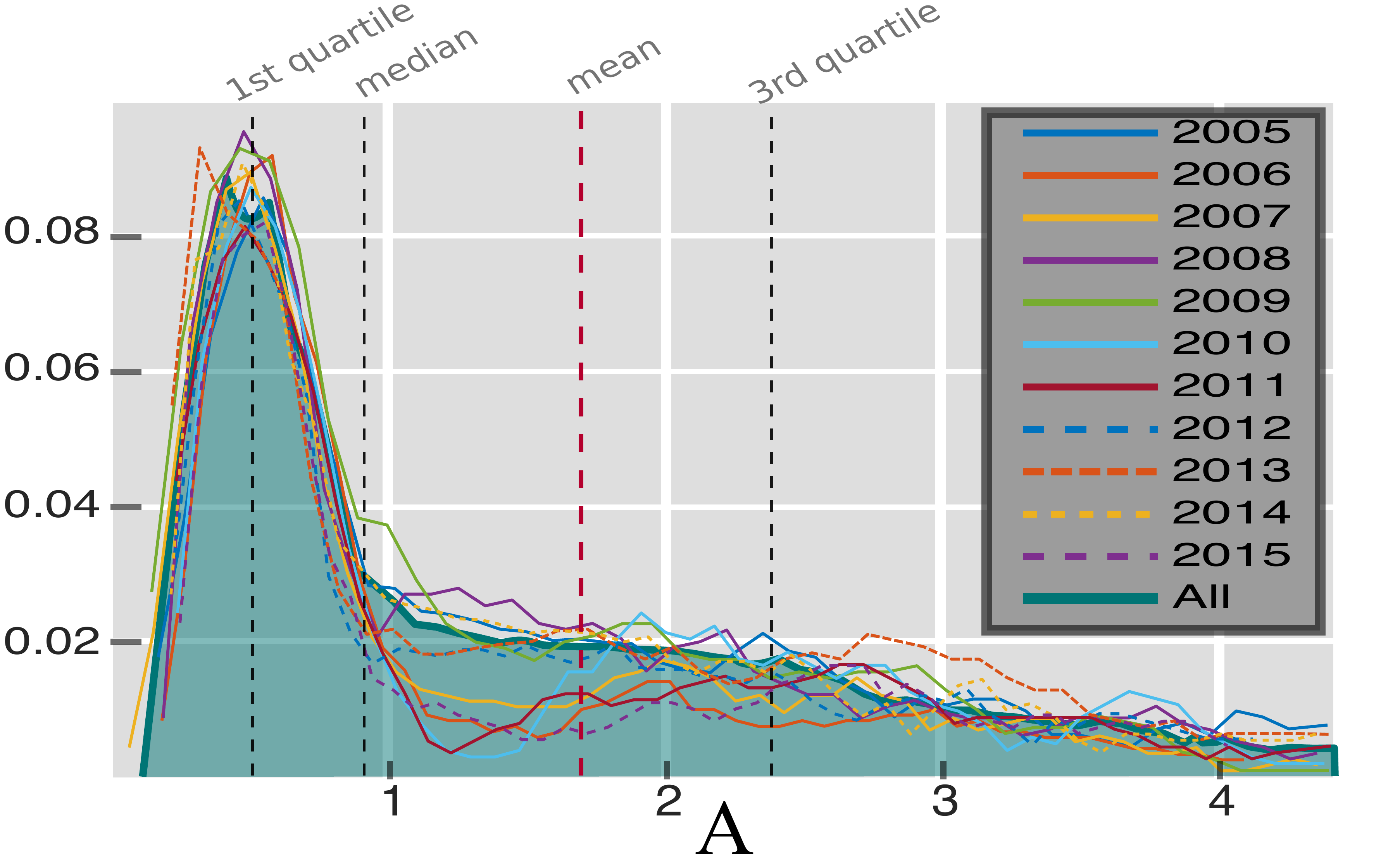}
                \caption{ }
                \label{A_hist_in_years}
        \end{subfigure}%
        \caption{ The histogram of $A$ (parameter of model as given by equation~\eqref{Gt}) for films as categorized by (a) genre, (b) month of release, (c) year of release.  }\label{A_hist}
\end{figure}

\begin{figure}[!h]
        \centering
        \begin{subfigure}[b]{0.3 \columnwidth}
                \includegraphics[width=\columnwidth,height=3cm]{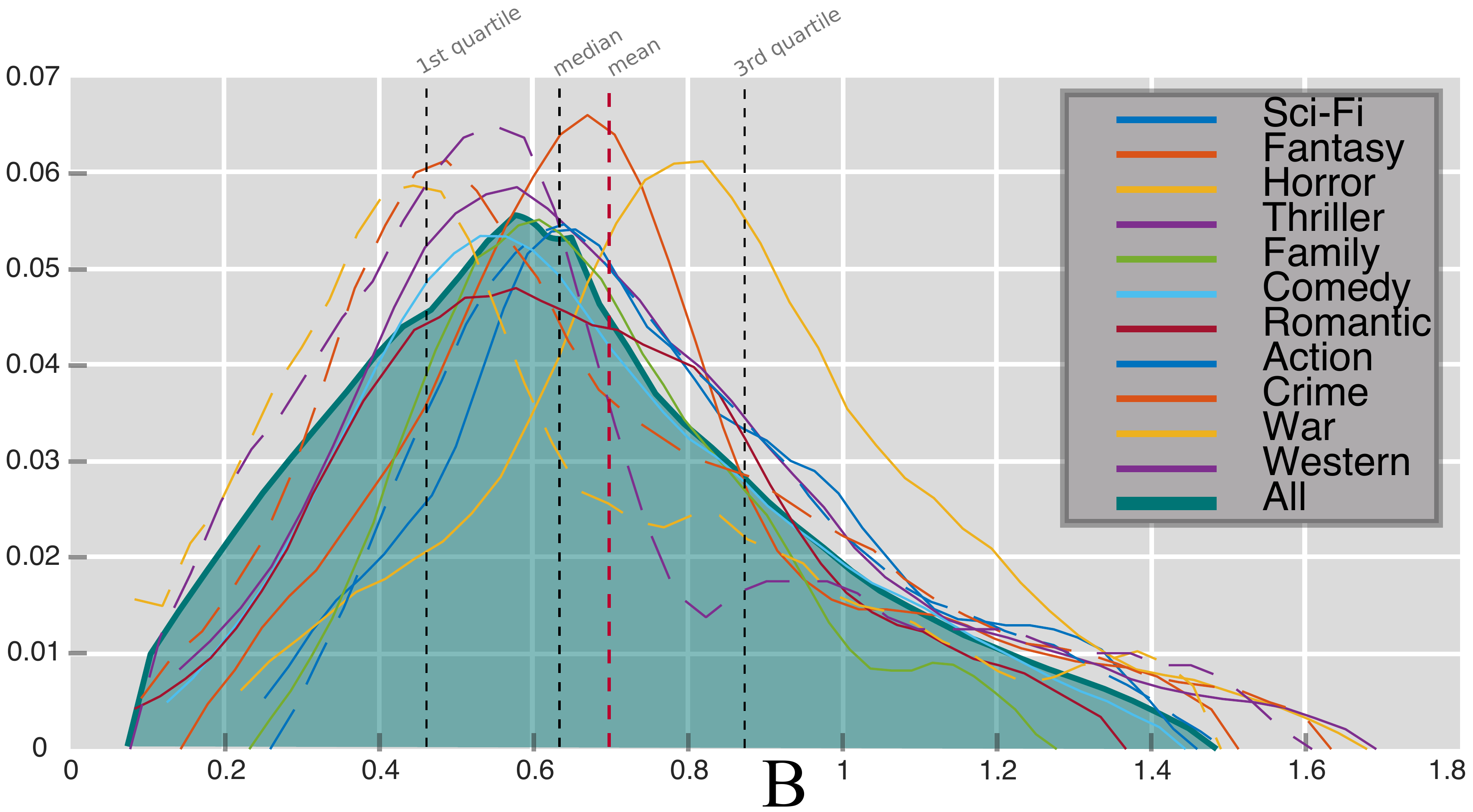}
                \caption{ }
                \label{B_hist_in_genres}
        \end{subfigure}%
        ~ 
            \begin{subfigure}[b]{0.3 \columnwidth}
                \includegraphics[width=\columnwidth,height=3cm]{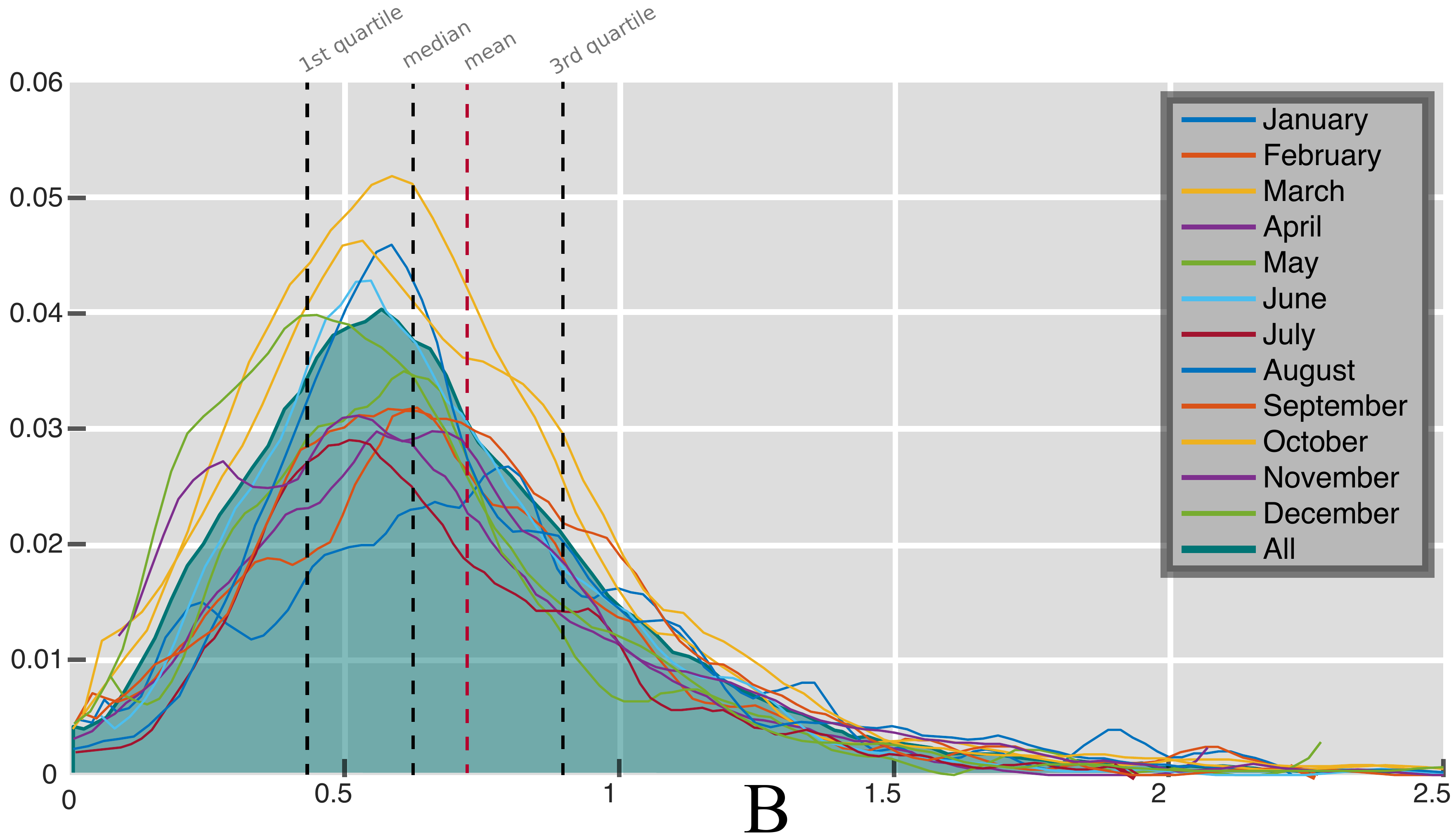}
                \caption{ }
                \label{B_hist_in_months}
        \end{subfigure}   
~
         \begin{subfigure}[b]{0.3 \columnwidth}
                \includegraphics[width=\columnwidth,height=3cm]{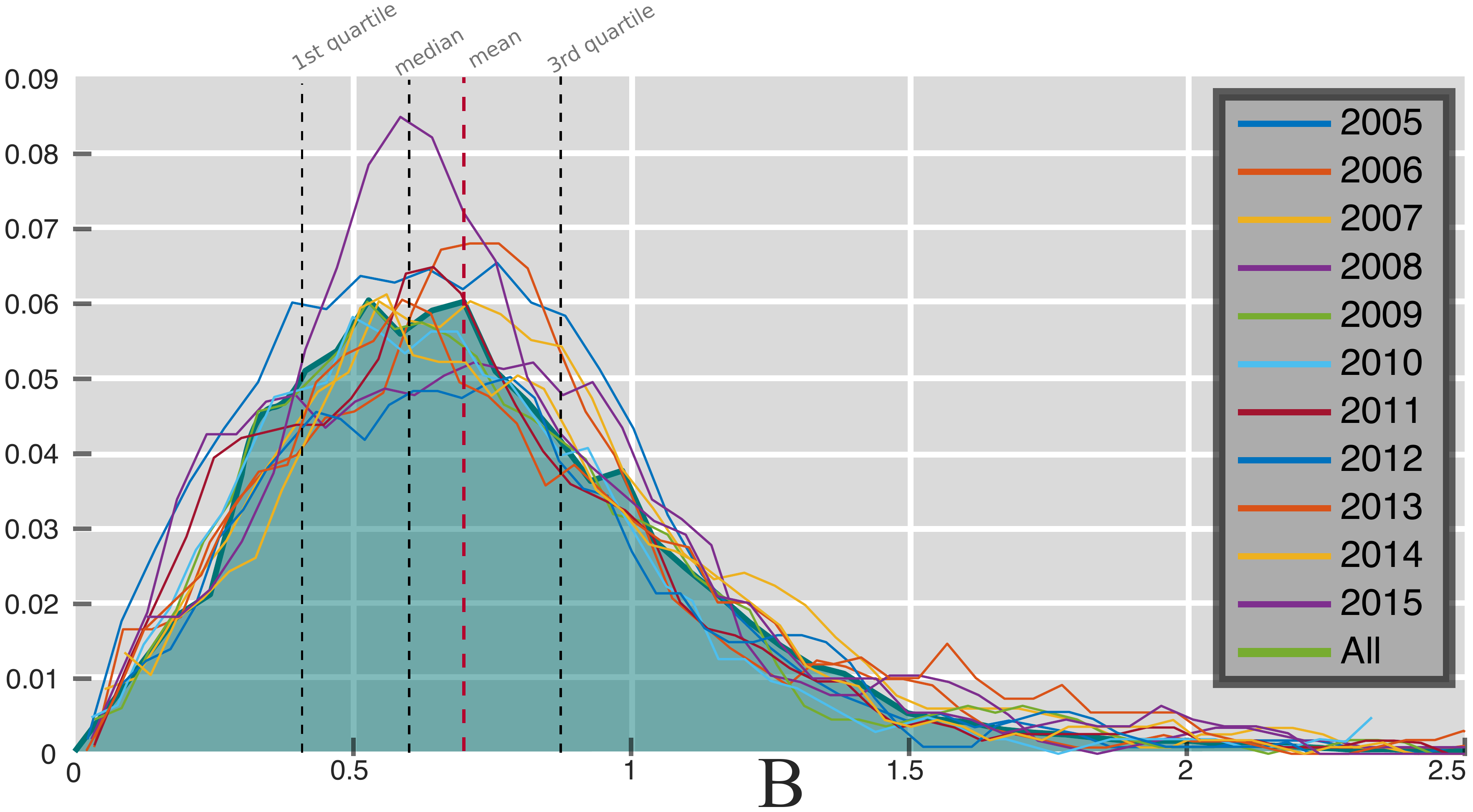}
                \caption{ }
                \label{B_hist_in_years}
        \end{subfigure}%
        \caption{ The histogram of $B$ (parameter of model as given by equation~\eqref{Gt}) for films as categorized by (a) genre, (b) month of release, (c) year of release.   }\label{B_hist}
\end{figure}

 But to make the proposed procedure more practicable for a broader readership and further emphasize the strength of the model,  we take an even  simpler approach.  
With an approximation, we transform the present problem into one of  simple linear regression. We achieve considerable simplicity at  the cost of losing some accuracy, which is reasonably small (as will be  demonstrated below).   Note that in the limit as $t \rightarrow \infty$, we have $z(t) \rightarrow A$. 
This means that if a  film would be allowed to run forever, $z(t)$ would approach $A$. As an approximation to infinite time, we take the up-to-last week  revenue of each film as an estimate of $A$. Denoting the lifetime of film by $L$, this means that we can estimate $A$ by $z(L)$. So we need to estimate $B$ in the 
equation $z(t)= z(L) (1-e^{-B t})$, which essentially means that with the transformation $y(t)\stackrel{\text{def}}{=} \log \big[ 1- z(t)/z(L) \big]$, we have an 
  simple linear regression problem in which $y(t)$ is a linear function of $t$, and $B$ can be readily estimated for each film. 
  The histogram of the regression $R^2$ for the 5000 films in the data set  is presented in Figure~\ref{R2} as stratified by (a) genre, (b) month of release, and (c) year of release. It can be readily seen that the increase in error in the said simplification has not been substantial.  
The histogram of estimates for $A$ is presented in Figure~\ref{A_hist} as stratified by (a) genre, (b) release month, (c) release year. 
The histogram for $B$ is  presented in Figure~\ref{B_hist} with similar stratification.

\begin{figure}[!h]
        \centering
        \begin{subfigure}[b]{0.47 \columnwidth}
                \includegraphics[width=\columnwidth,height=5cm ]{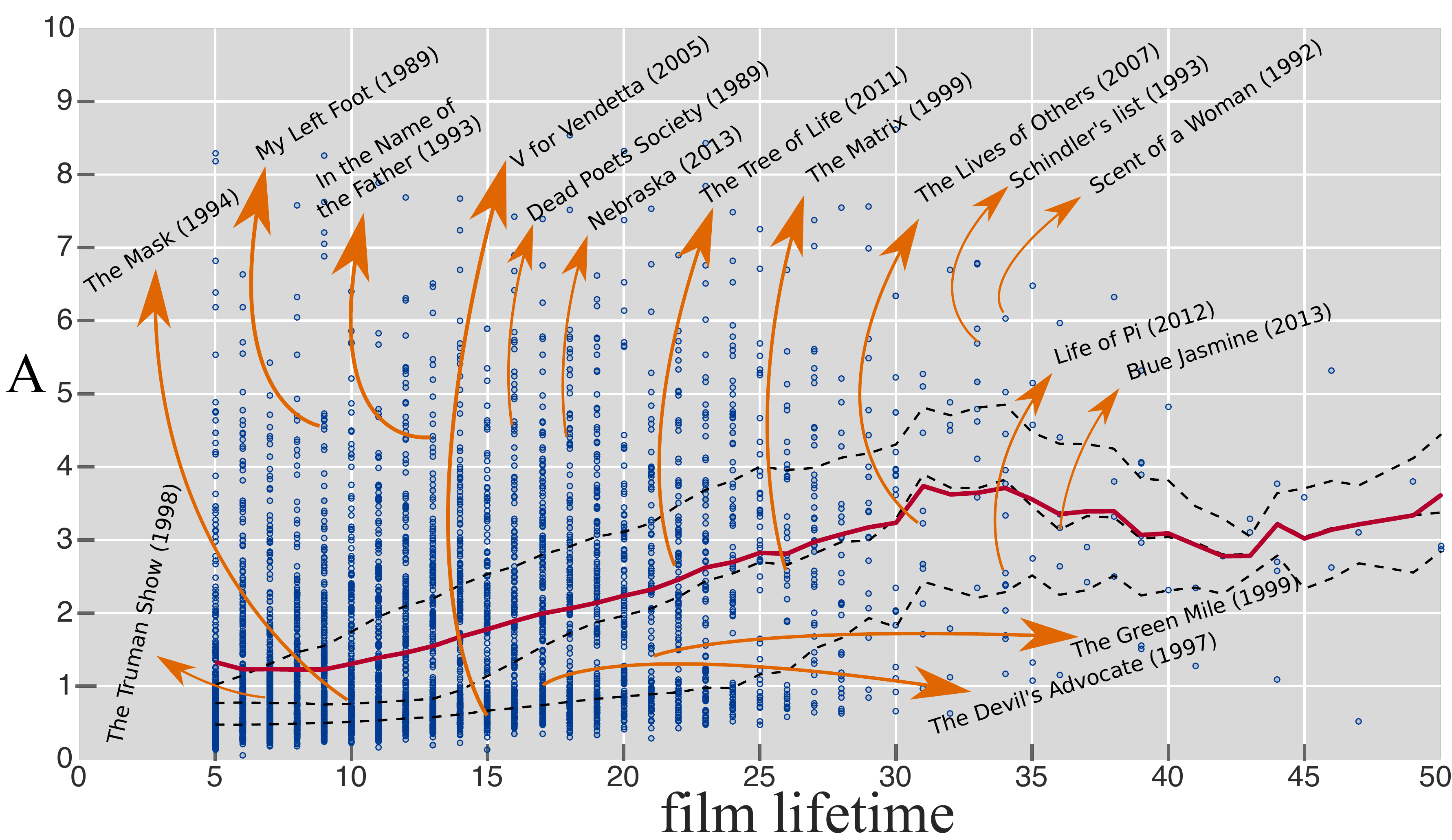}
                \caption{ }
                \label{A_in_lifetime}
        \end{subfigure}%
        ~ 
            \begin{subfigure}[b]{0.47 \columnwidth}
                \includegraphics[width=\columnwidth,height=5cm ]{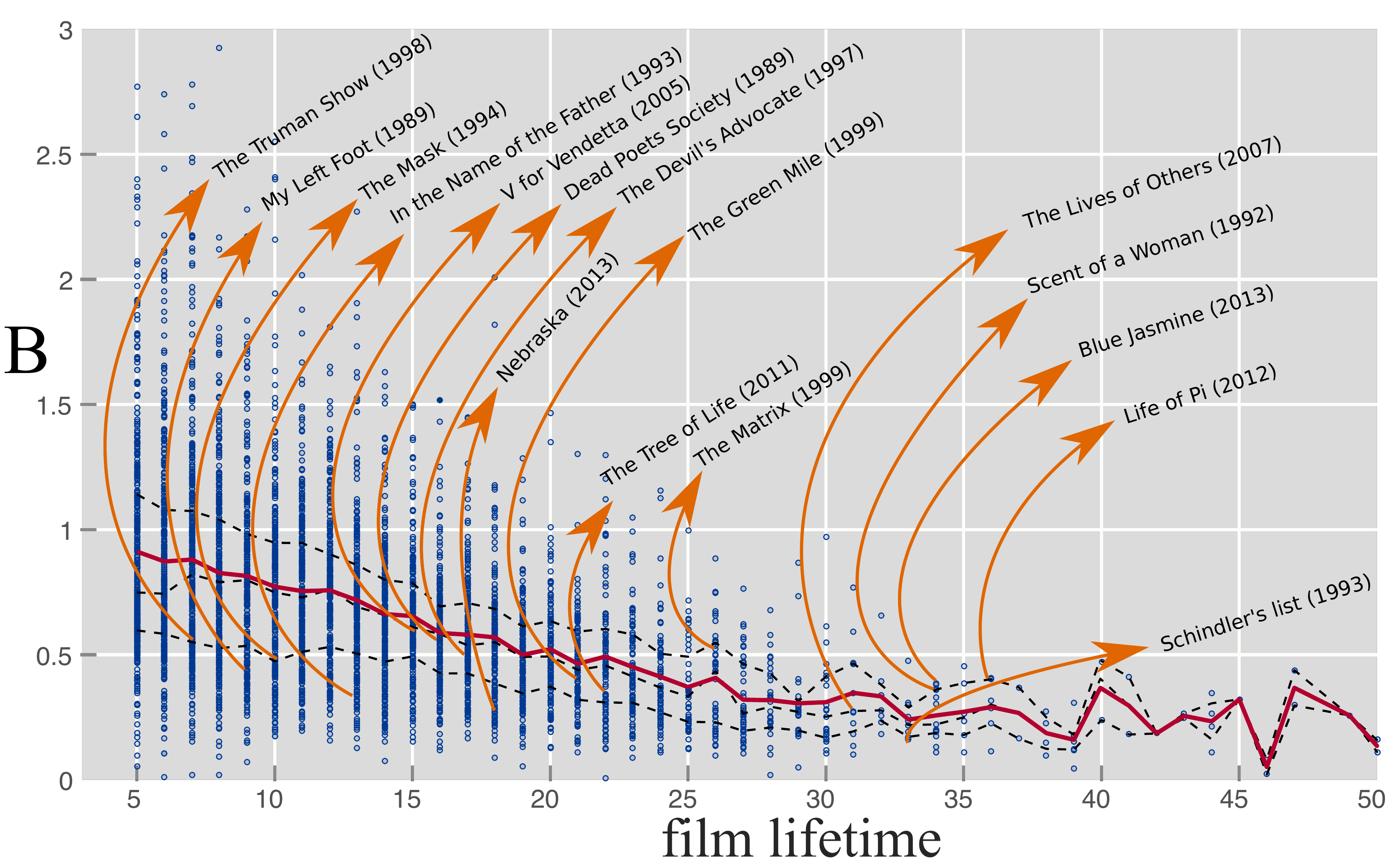}
                \caption{ }
                \label{B_in_lifetime}
        \end{subfigure}   
        \caption{ (a) and (b): Model parameters  as a function of film lifetime, and (c): A as a function of $\frac{1}{B}$. The red line pertains to the mean, and the  dashed lines pertain to the 25, 50, and 75 percentile. Some example films are pointed out  for illustrative purposes. 
        }\label{A_B_L}
\end{figure}

Figure~\ref{A_B_L} depicts  $A$ and $B$ as a function of    the number of weeks  the film remained in theaters (L). It can be seen that $A$ has a positive relationship with $L$, and that $B$ has a negative relationship, as expected (because increasing $A$---social influence---means increasing revenue, and increasing $B$ means faster memory decay).

\begin{figure}[!h]
        \centering
        \begin{subfigure}[b]{0.47 \columnwidth}
                \includegraphics[width=\columnwidth,height=4.8cm ]{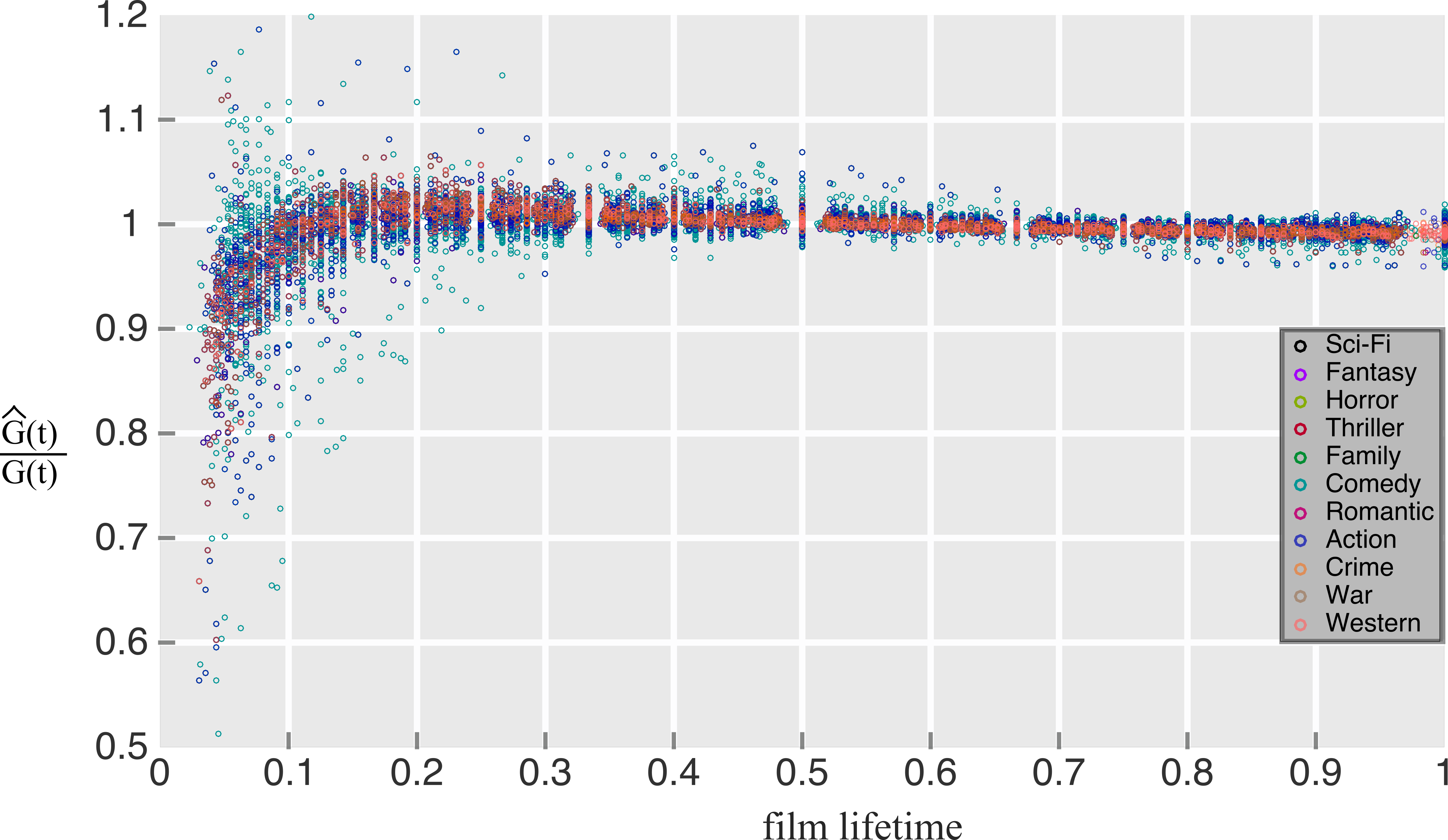}
                \caption{ }
                \label{accuracy_in_time}
        \end{subfigure}%
~
         \begin{subfigure}[b]{0.47 \columnwidth}
                \includegraphics[width=\columnwidth,height=5.1cm ]{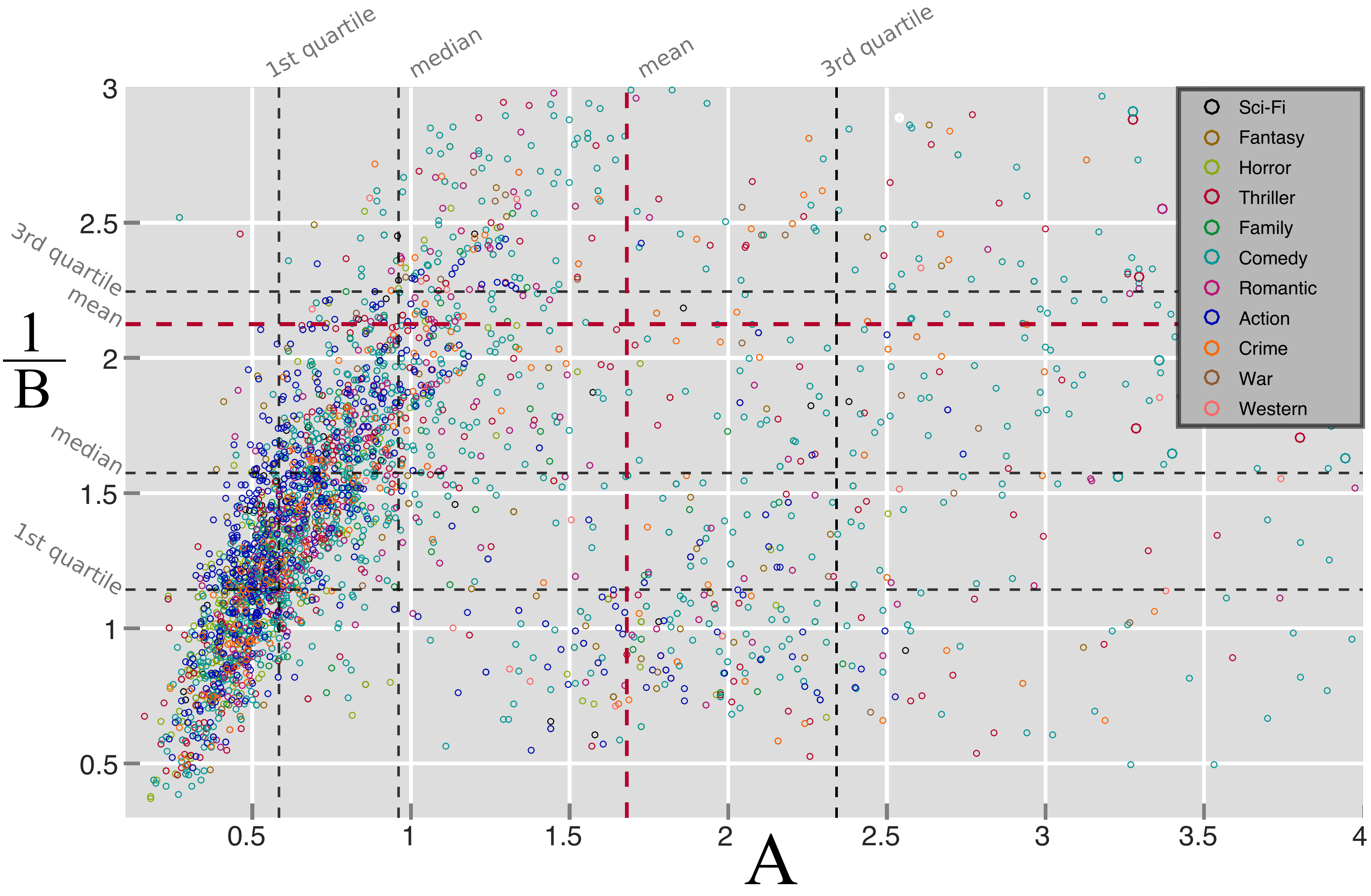}
                \caption{ }
                \label{A_vs_Binverse_good}
        \end{subfigure}%
        \caption{ (a)  the ratio of the predicted revenue over actual revenue for all films as a function of normalized time, that is, $t/L$. (b) A as a function of $\frac{1}{B}$
        }\label{accuracy}
\end{figure}

Denoting the estimated revenue from the above procedure by $\widehat{G}(t)$, we can plot $\widehat{G}(t)/G(t)$ to see how accurate the prediction is. Because of different lifetimes of different films (that is, different $L$ values due to different number of weeks films survive in theaters), we plot it as a function of normalized time, which is defined as $t/L$, so that we can  plot  all films at the same frame. Figure~\ref{accuracy_in_time} presents the results, stratified by genre.  Given the model simplicity, the errors are reasonably low. 
 Moreover, there is no visually discernible difference between the performance in different genres.

\begin{figure}[!h]
        \centering
        \begin{subfigure}[b]{0.47 \columnwidth}
                \includegraphics[width=\columnwidth,height=5.7cm]{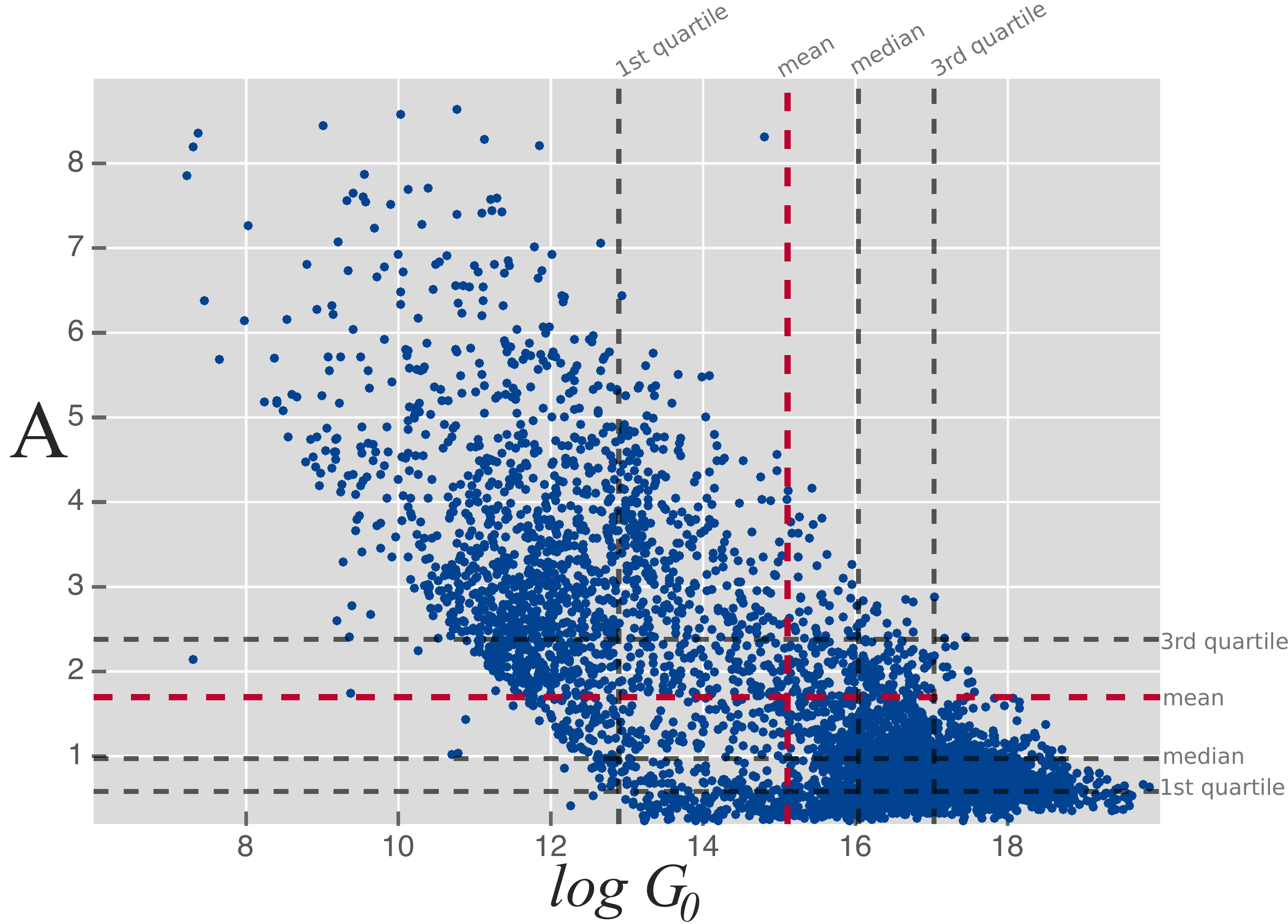}
                \caption{ }
                \label{A_vs_G0}
        \end{subfigure}%
        ~ 
            \begin{subfigure}[b]{0.47 \columnwidth}
                \includegraphics[width=\columnwidth,height=5.6cm]{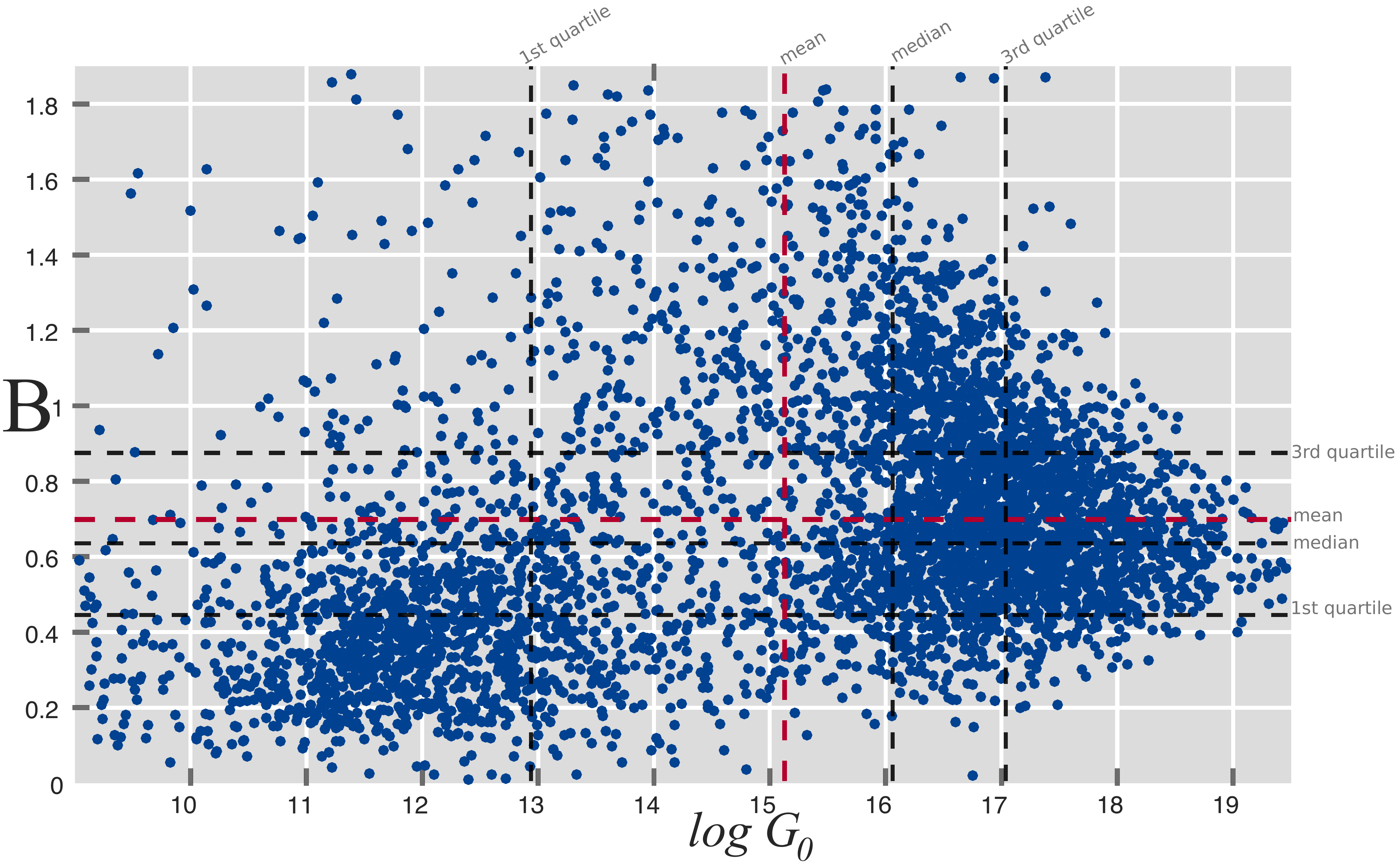}
                \caption{ }
                \label{B_vs_G0}
        \end{subfigure}%
        \caption{The parameters of the model as a function  of the revenue in opening week. The red line pertains to the mean, and the  dashed lines pertain to the 25, 50, and 75 percentile.   }\label{A_B_G0}
\end{figure}

 We also attempted to simplify the model even further by eliminating one of the parameters by well-approximating  it  as a function of  the other parameter, so that we would have an elegant  one-parameter model. Visually, we find that for a considerable  portion of the films, the relationship of $A$ and $1/B$ seem   close to linear, as depicted in figure~\ref{A_vs_Binverse_good}. The correlation of $A$ and $1/B$ is indeed high ($\rho=0.60$). But since the linear relationship seems to hold only for a certain portion of the population, which is indeed the majority but not a strong one, we refrain from complete replacement of one of the parameters in terms of the other one based on this relationship. 
 We also attempt such a simplification by investigating possible linkages between $A$ and $B$  to $G_0$ (the first-week film revenue). 
 Figure~\ref{A_B_G0} presents the scatter plots of $A$ and $B$ as a function of $G_0$. 
  There is a clear negative relationship between $A$ and $\log G_0$ ($\rho=-0.59$,  and $p<10^{-5}$ for the t-test of the linear-association hypothesis). 
  There is also a small positive association between $B$ and $\log G_0$  ($\rho=0.22$,  and  $p<10^{-5}$ for the t-test of the linear-association hypothesis). 
  The association is weak because $B$ is modeling human memory in our simple model,  so it is not expected to vary much with other parameters.
   

As we emphasized above, the main focus of this study is to devise a mechanistic model that links the micro mechanisms of social influence to 
macro observable phenomena, and sales prediction is not a primary goal. But as a byproduct, we can utilize the results in order to 
add a predictive dimension to the results. We have estimated a value  for the $A$ parameter and one for the $B$ parameter for each film. 
To obtain a formula that would function as a reasonable `average behavior' for every film, we can take two approaches. 
The first approach is to pool all the $A$  estimates together and all the $B$  estimates together, and take the medians as the parameters for the predictive model. 
The other is to first stratify by genre (or any other attribute, we take genre here as an illustrative example), and take the median of the distributions of parameters only for films in that genre to predict sales of films in each genre. 
 Figure~\ref{accuracy} presents the results in these two cases. It is seen that the pooling procedure produces better results than the genre-specific method. 
Also note that conventional cross-validation methods are not applicable here, because in estimating the parameters for each film, other films have not been used. In other words, our data units are time series here, not data points one would fit a curve to. So cross validation would simply act on the median-taking procedure, not the parameter-learning process.

\begin{figure}[!h]
        \centering
        \begin{subfigure}[b]{0.47 \columnwidth}
                \includegraphics[width=\columnwidth,height=5cm ]{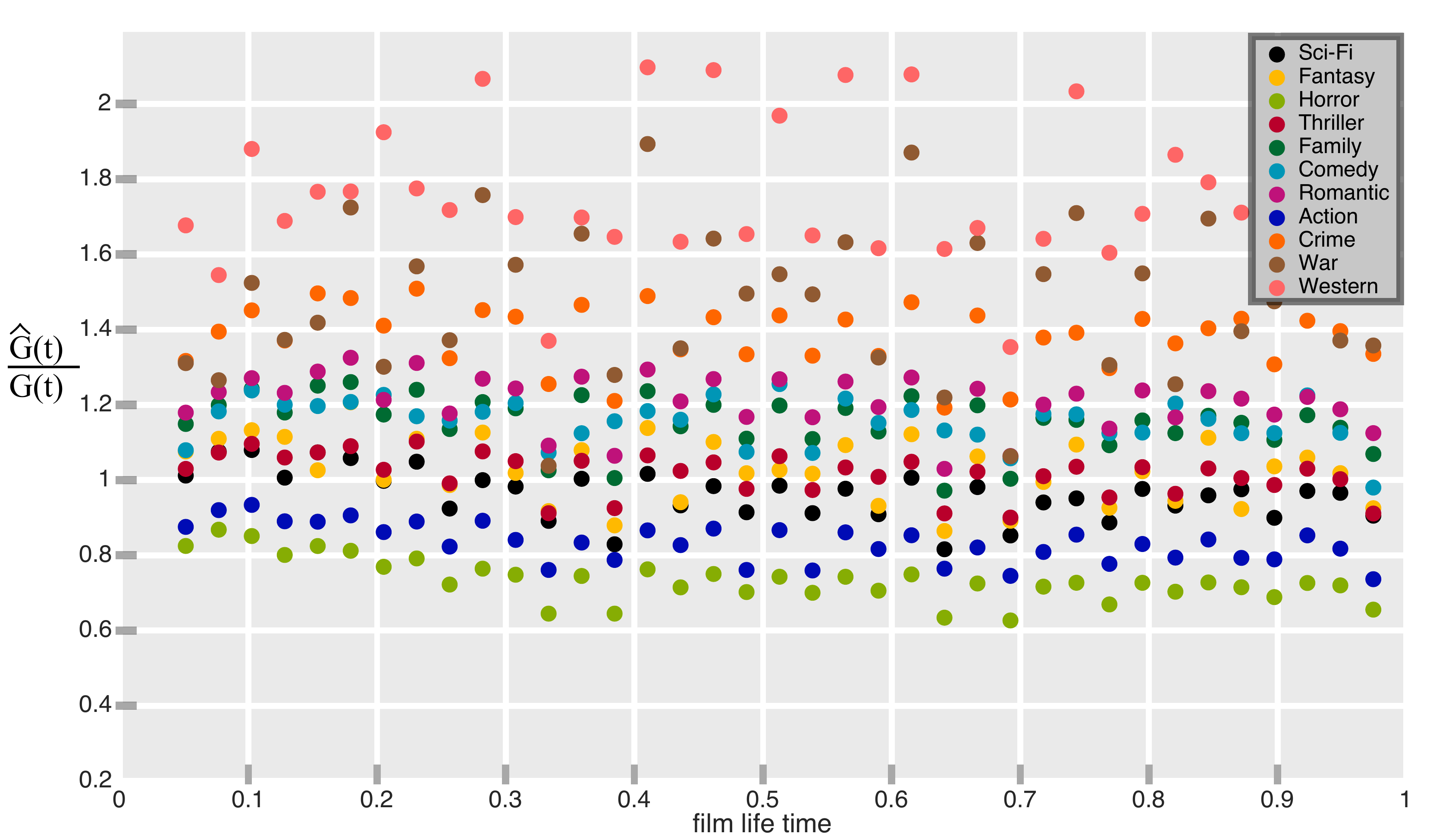}
                \caption{ }
                \label{pool_1}
        \end{subfigure}%
~
         \begin{subfigure}[b]{0.47 \columnwidth}
                \includegraphics[width=\columnwidth,height=5cm ]{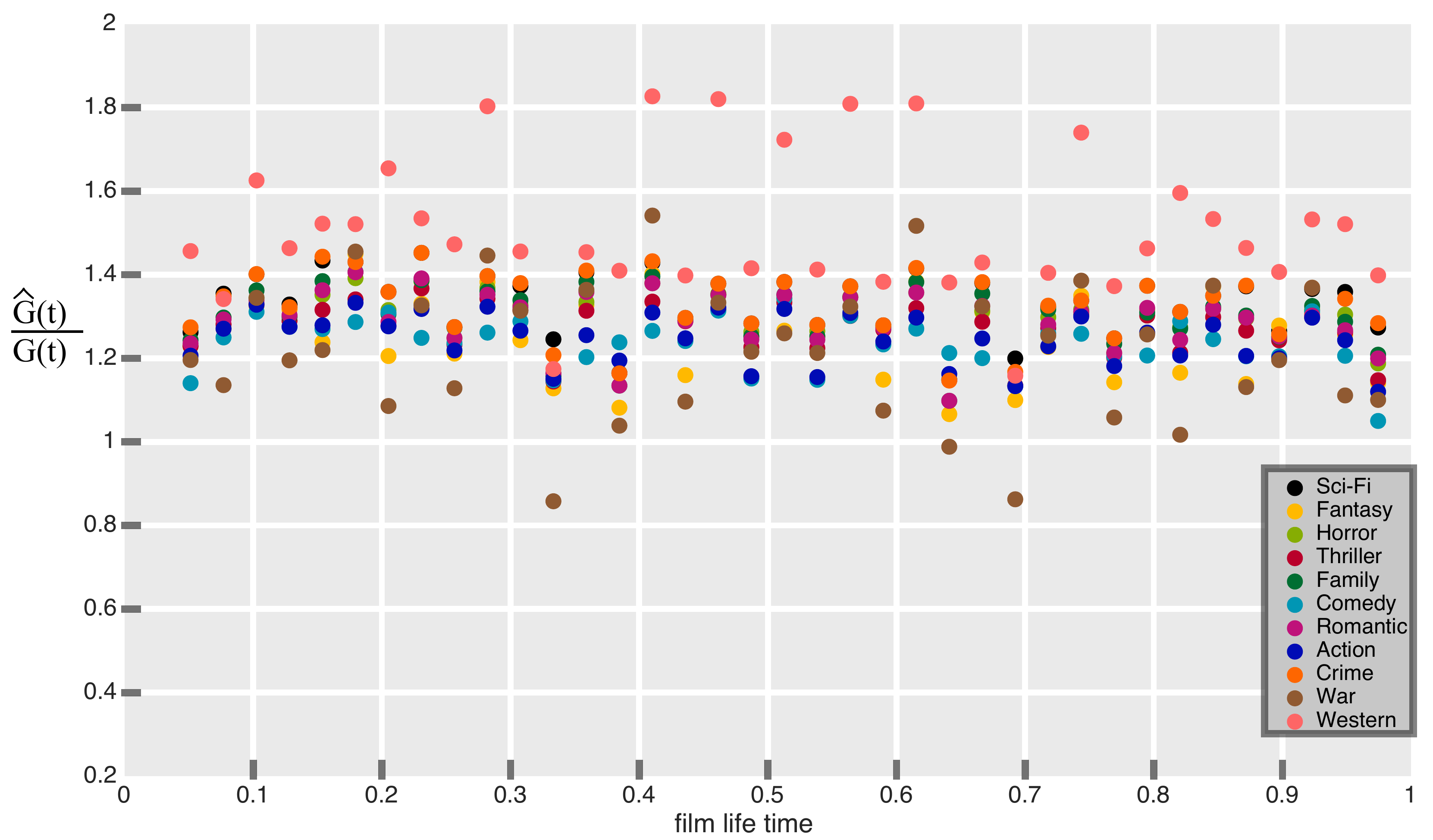}
                \caption{ }
                \label{pool_2}
        \end{subfigure}%
        \caption{ 
    Ratio of predicted value of $G(t)$ to the empirical value as a function of normalized film lifetime, using median values in the distributions of $A$ and $B$ across (a) all films pooled together, (b) only films of given genre.    
        }\label{accuracy}
\end{figure}

We also investigate if $G_0$ can be replaced by pre-release data. 
 As an illustrative example, we  use the Wikipedia page hits  ($W$) of films in the month prior to release as a proxy for $G_0$. As Figure~\ref{G0_wiki} presents, 
there is a  positive association between $\log G_0$ and $\log W$  (with $\rho=0.44$).  Accurate prediction of $G(0)$ from pre-release data 
has been already done in the literature with remarkable accuracy (e.g., $R^2>0.9$ using Wikipedia page edits and hits~\cite{Taha}, and  $R^2>0.95$ using Twitter mentions 
one day prior to release~\cite{asur2010predicting}). So for the reader interested primarily in prediction, there is an  avenue for improvement 
of prediction ability by combining pre-release data with the results presented in this paper.

\begin{figure}[!h]
        \centering
        \begin{subfigure}[b]{0.5 \columnwidth}
                \includegraphics[width=\columnwidth ]{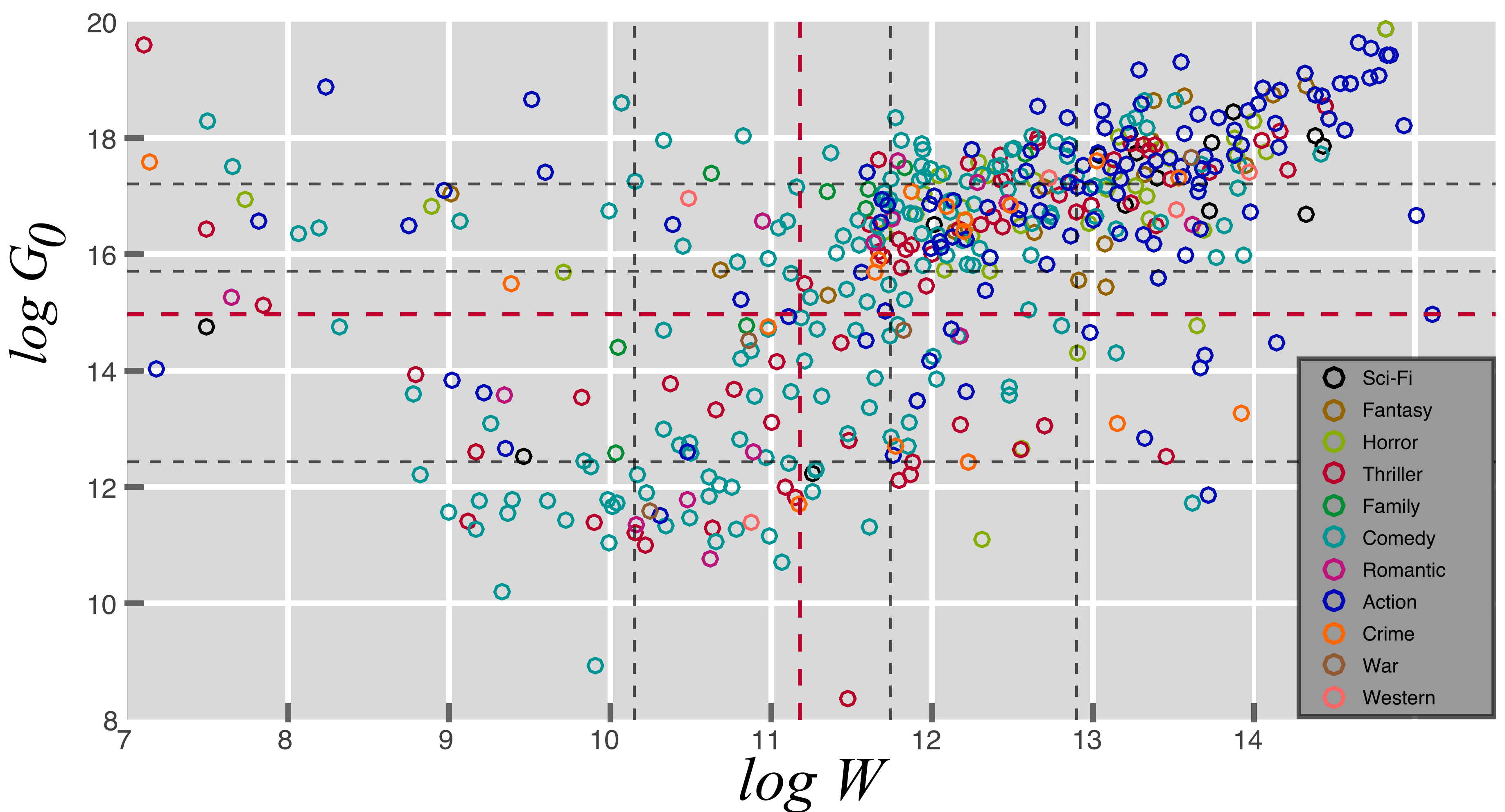}
                \caption{ }
                \label{G0_wiki}
        \end{subfigure}%
        \caption{The parameters of the model as a function  of the revenue in opening week  }\label{G0_wiki}
\end{figure}

\section*{Discussion}
The distribution of $A$ (presented in Figure~\ref{A_hist}) is skewed. 
It is known in the literature that film revenues have a heavy-tailed distribution~\cite{sinha2004hollywood,sinha2005blockbusters}. 
Since greater $A$ means greater social influence, and this in turn means more revenue (and longer lifetime in the theaters, as shown in Figure~\ref{A_in_lifetime}), we would expect that such a heavy-tailed distribution would be reflected in the distribution of $A$, too. 
As we discussed above, $B$  (that characterizes memory decay) models   intrinsic properties of the human memory system and 
is not expected to vary much across films. As Figure~\ref{B_hist} illustrates, $B$ is fairly localized, which does match this expectation. Unlike $A$, parameter $B$ is expected to have a negative relationship with revenue, and consequently, with $L$ (film lifetime). This is confirmed empirically, as illustrated in Figure~\ref{B_in_lifetime}). 
The expected  increase of $L$ with $A$ and its decrease with $B$ are both statistically significant (hypothesizing a  linear relationship, we have $p<10^{-5}$ for both t-tests).

   As illustrated in Figure~\ref{A_B_G0}, there  is a clear negative relationship between $A$ and $\log G_0$ ($\rho=-0.59$,  and $p<10^{-5}$ for the t-test of the linear-association hypothesis). This is expected because a higher $G_0$ means  a stronger role  for marketing campaigns and the direct influence individuals receive by encountering advertisement. For example,   multitudes of people were individually waiting for the release of the recent Star Wars: The Force Awakens (2016) movie, not waiting for the feedback of others to make a decision.
   
  There is also a small positive association between $B$ and $\log G_0$  ($\rho=0.22$,  and  $p<10^{-5}$ for the t-test of the linear-association hypothesis). 
   The causes of this positive association  are  purely structural. It does not mean that movies with stronger premier are forgotten faster. 
  It merely reflects the fact that  the cinema market capacity is limited.  That is, people are either `movie-goers' or not,  and sudden change of  attitudes towards the film industry is not very common, especially in a single film lifetime. So higher $G_0$ necessarily means that the film has to decay faster. For example, Finding Dory (2016) has opening week revenue 231M, and The Jungle Book (2016) has 130M, but both of them  sold  31M in the 4th  week.  It is in this sense that we say the  one with higher $G_0$ has to decay faster.

 The predictions of the aggregated model   are more accurate when all films are pooled (Figure~\ref{pool_1}) than when genre-specific pooling is used (Figure~\ref{pool_2}). 
 The accuracy differs for different genres. 
  It is of note that genres with similar content do exhibit similar performances. War is close to Western,   Romantic is close to Family as well as Comedy (primarily due to   Romantic-Comedy  films that bring these two genres close), and Sci-Fi is close to Fantasy. 
  Also, note that differences in accuracy  are  not due to different subpopulation sizes (that is, it is not the case that, for example, performance for Sci-Fi is good because there are  many  Sci-Fi films in the data set, and that of War  is poor because there are only few War films), because  when genre-specific pooling is used (Figure~\ref{pool_2}), the performances are still distinct. This means that, as one would intuitively expect,  there  exist inherent differences  between films of different genres, particularly with regard to our model. 

 Since the model presented in this paper is minimal and we insisted on model simplicity as the central motive of this study, 
 extensions to the present model can readily be made.  Which  direction yields the most improvement is an interesting question on its own. We ignored individual influence received from marketing campaigns, which can be added to the model with a new parameter. That way, there will be two distinct sources of influence. One would be social, as included here, and the other will be individual (through direct contact with advertisements in the media, billboards, etc.). The latter will add a new parameter to the model. One conjecture would be that  susceptibility to social influence would be highly correlated with susceptibility to direct marketing, so that the new parameter could be reasonably well-approximated as a function of already-existing parameters.
  Another improvement would be to use  data from   social media to  incorporate the effects of network position of individuals on their social influence, 
  and the effects of  structural properties of the underlying  social networks in the diffusion of decisions.

\section*{Methods}\label{methods}

\subsection*{Model Analysis} 
Let us denote the state of individual $x$ by $s_x$, which is zero if the individual is in the $S$ state and is one if the individual is in the $I$ state. 
 For a given $I$ social tie, the probability that after time $dt$ the state will be transmitted to individual $x$ is given by $\beta dt$, by definition. 
 The probability that transmission does not occur is ${1-\beta dt}$. The probability that the individual does not get infected from any of existing social contacts is $(1-\beta dt)^{i_x}$, where $i_x$ is the number of social contacts of individual $x$ who are in state $I$. Thus, the probability that individual $x$ will be infected after time $dt$ is 
 ${1-(1-\beta dt)^{i_x}}$, which can be Taylor-expanded to first order of $dt$ to $\beta i_x dt$. Under the mean-field approximation, the expected state of individual $x$ with $k_x$ total social contacts  after time $dt$ can be written as ${E\{s_x(t+dt) \} = 
s_x + (1-s_x) (\beta k_x \rho_t ) dt }$.  Averaging this over all individuals, we get ${\dot{\rho}_t= \rho_t (1-\rho_t) \beta  \langle k \rangle }$, 
where $\langle k \rangle $ is the average number of social contacts of individuals. Multiplying by the memory factor and integrating both sides, 
we get  Equation~\eqref{Gt}.

 Similar results can be derived for the    voter-type alternative model discussed in the text. In this case, we have 
 ${E\{s_x(t+dt) \} = 
s_x + (1-s_x)\alpha dt  ({I_x}/{k_x}) e^{-Bt}
}$, 
and if we sum this for all nodes, we get ${\dot{\rho}_t= \rho_t (1-\rho_t)  \alpha \sum_{xy} A_{xy}/kx }$, where $A_{xy}$ is the adjacency matrix of the underlying 
social network (it equals 1 if $x$ is connected to $y$, and equals zero otherwise). Multiplying by the memory factor, the 
product of  the two  coefficients ($\alpha$ which pertains to social influence and  $ {\sum_{xy} A_{xy}/kx }$ which pertains to social connectivity) can be absorbed into 
a single new parameter, and after integration we arrive at a result identical to the one for the epidemic model.

\subsection*{Data Description}

 \paragraph{Film Data}
 We extracted the publicly-available data set from \url{www.boxofficemojo.com}. We limited the analysis       to the 5000 top US-grossing films. 
 We excluded films before 1980 o ensure data reliability. We also excluded imax films which are not conventional feature films, and their lifetime is unusually long. For example, Space Station 3-D  imax was released in 2002, and is still being screened. We limit the analysis only to `traditional' films. We applied a cutoff of 70 weeks and   omitted  films with lifetimes longer than 70 weeks. Also, we excluded films with lifetimes shorter than 5 weeks. 
 We also excluded  the films released in 2016   to ensure  that no film in the data set is still in theaters. 
  The cleaned data set with sales time series accompanies this paper. 

 \paragraph{Wikipedia Data}
   We extracted month-of-release Wikipedia page hits from  the publicly-available data set provided by the Wikimedia foundation. 
   Although the page hits data set goes back to 2007,   we excluded films released prior to 2010 to make sure we are considering the period in which Wikipedia has become publicly established and popular enough to be a reliable source of data.



%

\section*{Author contributions statement}

N.M and B.F. formulated the problem.
 A.T., N.M., and B.F. performed the research. 
 A.T., N.M., B.F., and M.R. discussed the results and contributed to the text.


 \section{Competing Financial Interests Statement}
 The authors declare no competing financial interests.

\end{document}